\documentclass{article}

\usepackage{arxiv}

\usepackage[utf8]{inputenc} 
\usepackage[T1]{fontenc}    
\usepackage{hyperref}       
\usepackage{url}            
\usepackage{booktabs}       
\usepackage{amsfonts}       
\usepackage{nicefrac}       
\usepackage{microtype}      
\usepackage{lipsum}		
\usepackage{graphicx}
\usepackage{doi}
\usepackage{graphicx}
\usepackage{dcolumn}
\usepackage{bm}

\usepackage[utf8]{inputenc}
\usepackage[T1]{fontenc}
\usepackage{mathptmx}
\usepackage{etoolbox}
\usepackage{hyperref}
\usepackage{soul}
\newcommand{\mathcolorbox}[2]{\colorbox{#1}{$\displaystyle #2$}}
\usepackage[dvipsnames]{xcolor}
\usepackage{amsmath}

\title{Semi-supervised physics guided deep learning framework for predicting the I-V characteristics of GAN HEMT}

\author{\href{https://orcid.org/0000-0002-4015-4181}{\includegraphics[scale=0.06]{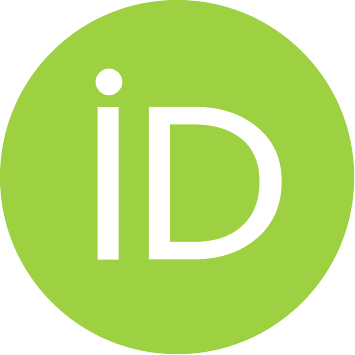}\hspace{1mm}Shivanshu Mishra} \\
	Central Electronics Engineering	Research Institute\\
	Pilani, India, 333031\\
	Academy of Scientific and Innovative Research\\
	Ghaziabad, India, 201002\\
	\texttt{shivanshuceeri@gmail.com} \\
	\And
	{\href{https://orcid.org/0000-0003-1605-7303}{\includegraphics[scale=0.06]{orcid.pdf}}\hspace{1mm}Bipin Gaikwad} \\
	Central Electronics Engineering	Research Institute\\
	Pilani, India, 333031\\
	Academy of Scientific and Innovative Research\\
	Ghaziabad, India, 201002\\
	\And
	{\hspace{1mm}Nidhi Chaturvedi} \\
	Central Electronics Engineering	Research Institute\\
	Pilani, India, 333031\\
	Academy of Scientific and Innovative Research\\
	Ghaziabad, India, 201002\\
	
}

\renewcommand{\shorttitle}{\textit{arXiv} Template}

\hypersetup{
pdftitle={A template for the arxiv style},
pdfsubject={q-bio.NC, q-bio.QM},
pdfauthor={David S.~Hippocampus, Elias D.~Striatum},
pdfkeywords={First keyword, Second keyword, More},
}

\begin{document}
\maketitle

\begin{abstract}
This letter proposes a novel deep learning framework (DLF) that addresses two major hurdles in the adoption of deep learning techniques for solving physics-based problems: 1) requirement of the large dataset for training the DL model, 2) consistency of the DL model with the physics of the phenomenon. The framework is generic in nature and can be applied to model a phenomenon from other fields of research too as long as its behaviour is known. To demonstrate the technique, a semi-supervised physics guided neural network (SPGNN) has been developed that predicts I-V characteristics of a gallium nitride-based high electron mobility transistor (GaN HEMT). A two-stage training method is proposed, where in the first stage, the DL model is trained via the unsupervised learning method using the I-V equations of a field-effect transistor as a loss function of the model that incorporates physical behaviors in the DL model and in the second stage, the DL model has been fine-tuned with a very small set of experimental data. The SPGNN significantly reduces the requirement of the training data by more than 80\% for achieving similar or better performance than a traditional neural network (TNN) even for unseen conditions. The SPGNN predicts 32.4\% of the unseen test data with less than 1\% of error and only 0.4\% of the unseen test data with more than 10\% of error.
\end{abstract}

\

Gallium nitride high electron mobility transistors (GaN HEMTs) have emerged as excellent devices as it provides best of both the material (GaN) and heterostructure advantages.
Superior material and electrical properties of the combination have led to a wide spectrum of applications in power amplifiers [1], power converters [2], high-frequency switches [3], high-temperature operations [4], and recently in sensors [5,6] and biosensors [7,8]. To achieve the aforementioned objectives and quality standards, the development of an accurate, fast, efficient and compact model is an important area of research.  Various efforts have been dedicated to the development of the GaN-HEMT models; the modelling approach can be majorly classified into three categories i.e., physics and semi physics-based, empirical modelling and numerical solution [9,10]. The first two approaches are generally a simple form of complex nonlinear behaviour that generates biases in the model that reduce the robustness and accuracy of the model. The numerical modelling provides good accuracy but it requires detailed information regarding the internal structure, device geometry, and material properties of the GaN-HEMT and uses complex simulation tools such as technology computer-aided design (TCAD) tools. It is both financially as well as computationally expensive and requires a lot of development time, execution time and expertise to obtain the output.

In recent years, the adoption of DL models for advancing the modelling approaches to solve physics-based problems has started due to their capability of finding patterns and complex relationships in given data. The DL models have already outperformed physics-based models in various disciplines such as materials science [11], applied physics [12], atmospheric science [13], chemical science[14], biomedical science [15] and computational biology [16]. However, the adoption of DL techniques in the field of semiconductors is very limited. The biggest hurdle that limits the application of DL models in physics-based problems is the scarcity of a large amount of quality data during the training phase. As the DL model is trained solely on the data, it does not incorporate the understanding of physics that sometimes leads to physically inconsistent results [17]. Physics informed or physics-guided neural network (PINN or PGNN) is an emerging area in the field of DL research that attempts to provide physics consistent results. The first paper on PGNN is published on the classification of miscible and immiscible systems of binary alloys in August 2017 by Zhang et.al [18]. In PGNN, the loss function of the DL model is modified to enforce physical constraints for e.g., Karpatne et.al, incorporated a monotonic decreasing function of water density and depth of the lake [19]. Similarly, the adoption of PGNNs has been increased across various fields [20-23]. However, modifying the loss function slightly depending on the nature of the study does not completely solve the problem of physically inconsistent results. Besides that, the existing PGNN works require a huge amount of labelled data which automatically creates a huge barrier for the research area where having a large amount of experimental data is not feasible. Thus, the existing techniques of PGNN try to incorporate a small extent of physics only. Also, the requirement of the huge amount of labelled data to train a DL model is still an untouched problem so far.

In this study, we have developed a novel semi-supervised deep learning framework (DLF) for the prediction of I-V characteristics of a GaN HEMT that significantly reduces the data required for training a DL model. The framework is generic in nature and can be applied to model a phenomenon from other fields of research too as long as its behaviour is known. The DLF can also act as an equation solver. In the proposed DLF, the DL model  is trained without data (unsupervised learning) through a physics based loss function in the first phase and later fine-tuned on a small dataset for accurate prediction in the second phase. The incorporation of the physics based loss function enables the DLF to predict the output values as good as the experimental results with significantly reduced training data and computational resources in real-time. The experimental evaluations are implemented in Python programming language (v3.6.6) using PyTorch library (v1.9.0).

The DL model consists of an input layer, 4 hidden dense layers and an output layer for the prediction of drain current of GaN HEMTs. The 4 hidden layers have neurons of 10, 10, 20 and 20 units in each layer, respectively. The input layer consists of a 9-dimensional vector consisting of mobility ($\mu_n$), permittivity of AlGaN ($\epsilon_{AlGaN}$), thickness of AlGaN (t$_{AlGaN}$), unit gate width (UGW), gate length (L$_g$), distance from source to drain (L$_{sd}$), gate to source voltage (V$_{gs}$), threshold voltage (V$_t$), and drain to source voltage (V$_{ds}$). The detail of the implemented DL model is given in the mathematical compact form in equation \ref{Eq1:DLModel}.

\begin{equation}\label{Eq1:DLModel}
O_{(L,M)} = \phi\left(\sum_{i=1}^{N_{(L-1)}}[w_{M,i}^{[L]}*O_{(L-1),i}]+b_{M}^{[L]}\right)
\end{equation}
N$_0$=9, N$_1$=10, N$_2$=10, N$_3$=20, N$_4$=20, N$_5$=1 \linebreak
L: 1$\,\to\,$5, 1 and M: 1$\,\to\,$N$_L$, 1

\noindent where, L, N$_L$, O and $\phi$,  denote layer number, number of neurons in a particular layer, output of a particular layer and activation function of each neuron respectively. O$_0$, O$_1$ to O$_4$, and O$_5$ are input (9-dimensional vector),  hidden layers, and output (drain current) of the DL model respectively. $\phi$ is the rectified linear unit (RELU) and linear unit for neurons of hidden and output layers respectively. The w and b are weights and biases of a neuron in a layer. The loss function that facilitates learning of the model without data is given in equation \ref{Eq1:DLModel4}.

\begin{equation}\label{Eq1:DLModel4}
Loss= \frac{1}{B}\left[I_{ds,pred(i)} - \mathcolorbox{Apricot}{\left\{\begin{array}{lr}
        K^{'}.{\Delta V_{i}.V_{ds(i)} - 0.5.V_{ds(i)}^2}, & \Delta V_{i} > V_{ds(i)}\\
        K^{'}.0.5. \Delta V_{i}^2, & \Delta V_{i} \leq V_{ds(i)}
        \end{array}\right\}}\right]^2
\end{equation}
where,
\begin{equation*}\label{Eq1:DLModel155}
\Delta V_{i}=V_{gs(i)} - V_{ds(i)}
\;\;and\;\;
K^{'}=\frac{\mu_{n(i)} .E_{AlGaN(i)} .W}{T_{AlGaN(i)} .L_{g}.K}
\end{equation*}

\noindent The highlighted portion (apricot color) of the loss function is basic equation of a FET in linear and saturation region. This analytical equation in loss function incorporates the physics of the FET in the DL model and force the network to produce results that are consistent with physics of the FET.

\begin{figure}[ht]
\centering
\includegraphics[width=4 in]{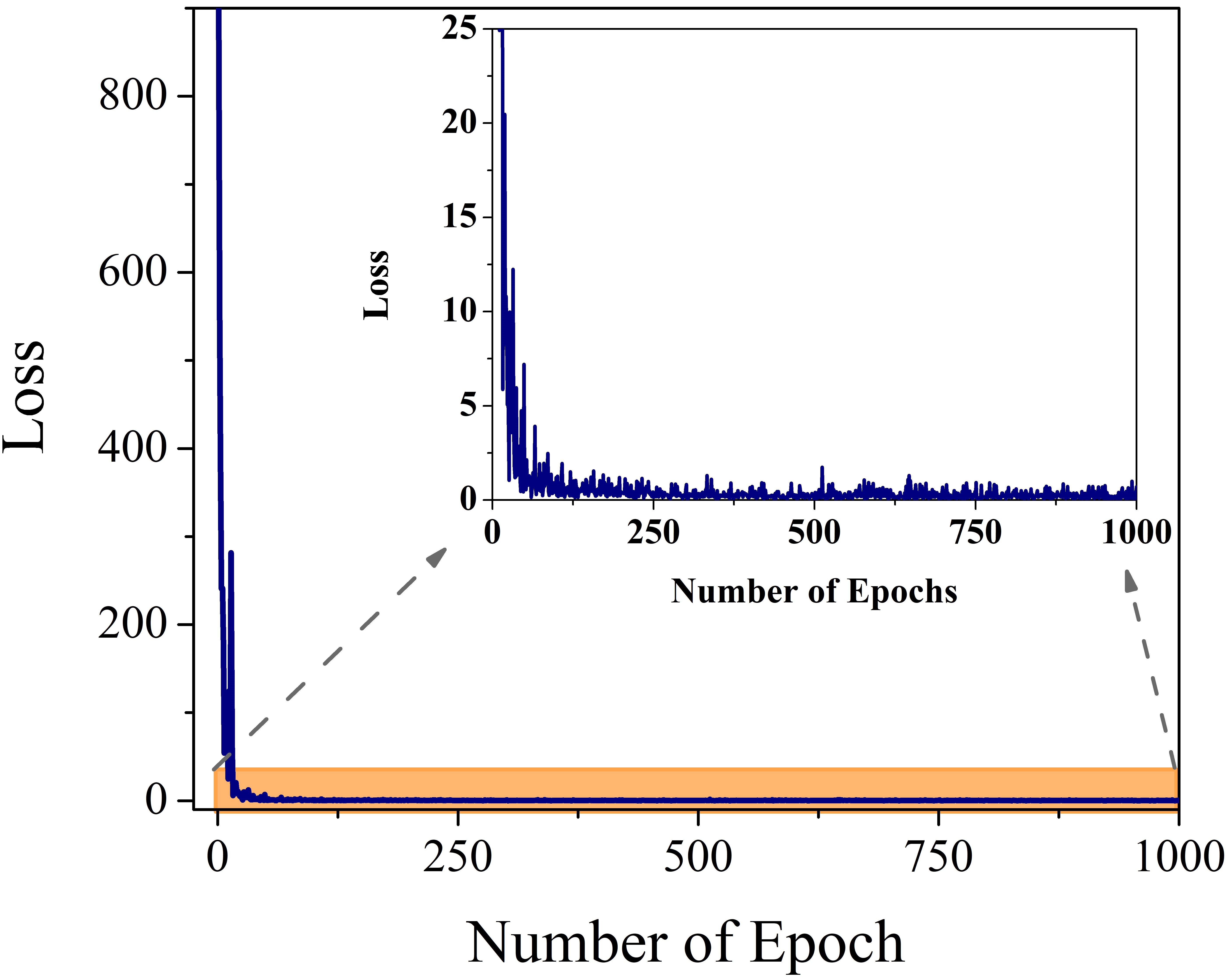}
\caption{Customized loss of the DL model based on the analytical I-V equation of the FET as function of each epoch.}
\label{fig:PreTrainLoss}
\end{figure}

The objective of the DL model in the first stage of training is to minimize the loss function that is given in equation \ref{Eq1:DLModel444444} 
\begin{equation}\label{Eq1:DLModel444444}
obj. =  \sum_{epochs}minimize (Loss)
\end{equation}

\noindent where epochs are the number of training iterations. An adaptive moment estimation has been employed to optimize equation \ref{Eq1:DLModel444444} with an initial learning rate of 5e$^{-3}$ that decays with a factor of 0.8 up to 1e$^{-5}$ if the plateau is observed for continuous 10 epochs. The DL model is initialized with weights and biases randomly taken from Glorot uniform distribution and associated loss has been backpropagated iteratively to update the parameters of the model to minimize loss value. After few epochs of training, the loss function starts decreasing and the DL model starts learning the I-V characteristics equations as shown in fig. \ref{fig:PreTrainLoss}. The model is trained for 1000 epochs and it completely learns the behaviour of the FET. The trained model itself becomes an equation solver that can instantaneously solve the FET equations.

\begin{figure}[ht]
\centering
\includegraphics[width=5in]{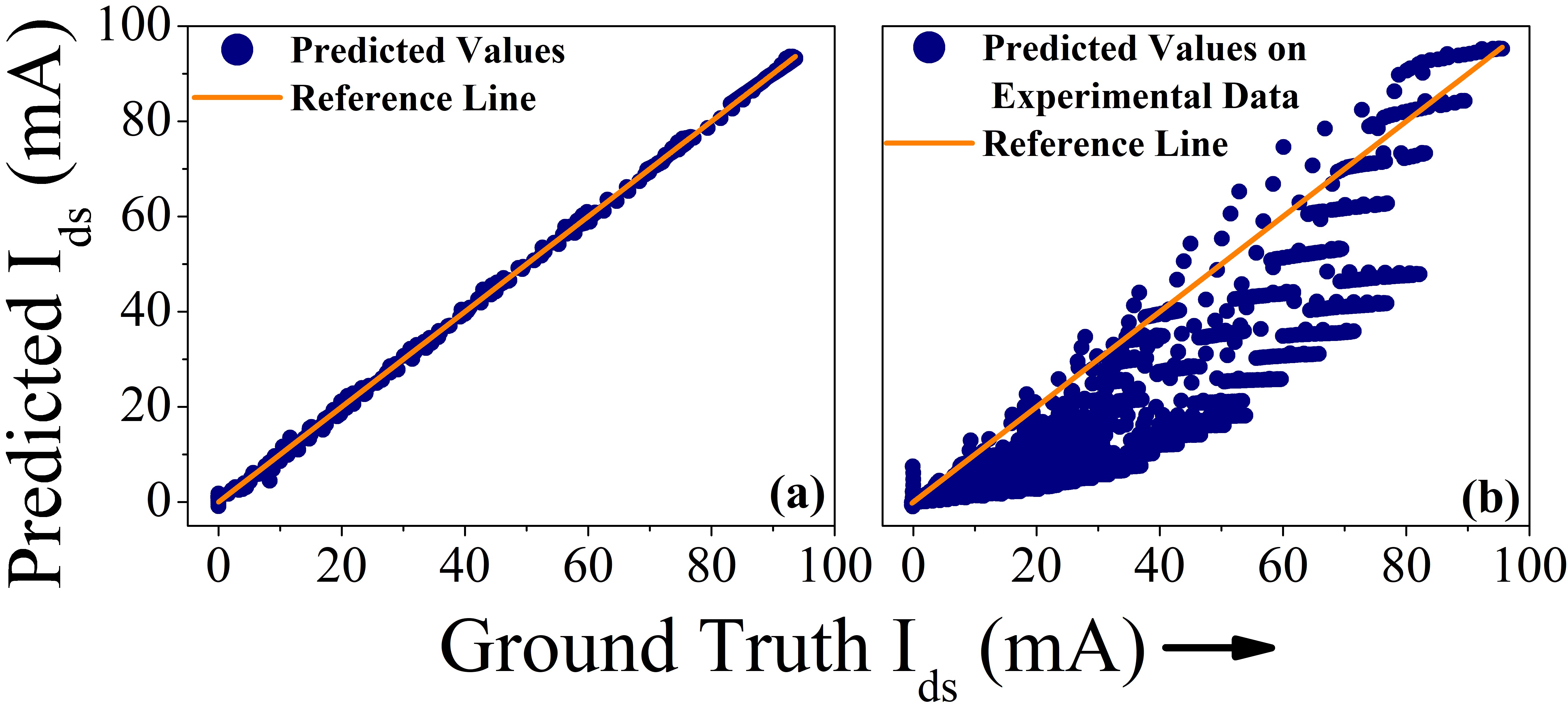}
\caption{Drain current predicted by the DL model (pre-train) trained in the first phase for test dataset of (a) analytical equation and (b) fabricated GaN HEMTs.}
\label{fig:PreTrainTvsP}
\end{figure}

The performance of the trained DL model has been compared with the I-V equation and experimental data as shown in fig. \ref{fig:PreTrainTvsP}. The associated experimental data of the fabricated GaN HEMTs of various designs have been previously published [6][24-26]. The model performs very well for solving the equations as shown in fig. \ref{fig:PreTrainTvsP}(a). However, for experimental data the performance deteriorates severely because the provided equations are very basic in nature and does not capture many other important factors. Now, by incorporating more complex effects in equation (2), the performance of the model can be enhanced significantly for experimental data as well similar to physics-based modelling and the neural network can work as a stand-alone equation solver. However, such modelling leads to very complex equations that might fit to the experimental data to some extent and often very difficult to obtain. Thus, our aim is to take this approach to a next level where a robust model that fits reasonably well to the experimental data can be developed with little exposure to FET physics and very small experimental data. A two-stage training approach has been utilized to optimize the parameters of the DL model. In the first stage, the DL model learns the the basic physics of the FET via unsupervised learning with physics-based loss function. Later, the pre-trained DL model from the first stage is fine-tuned on a small set of experimental data through supervised learning in the second stage.

\begin{figure}[ht]
\centering
\includegraphics[width=5 in]{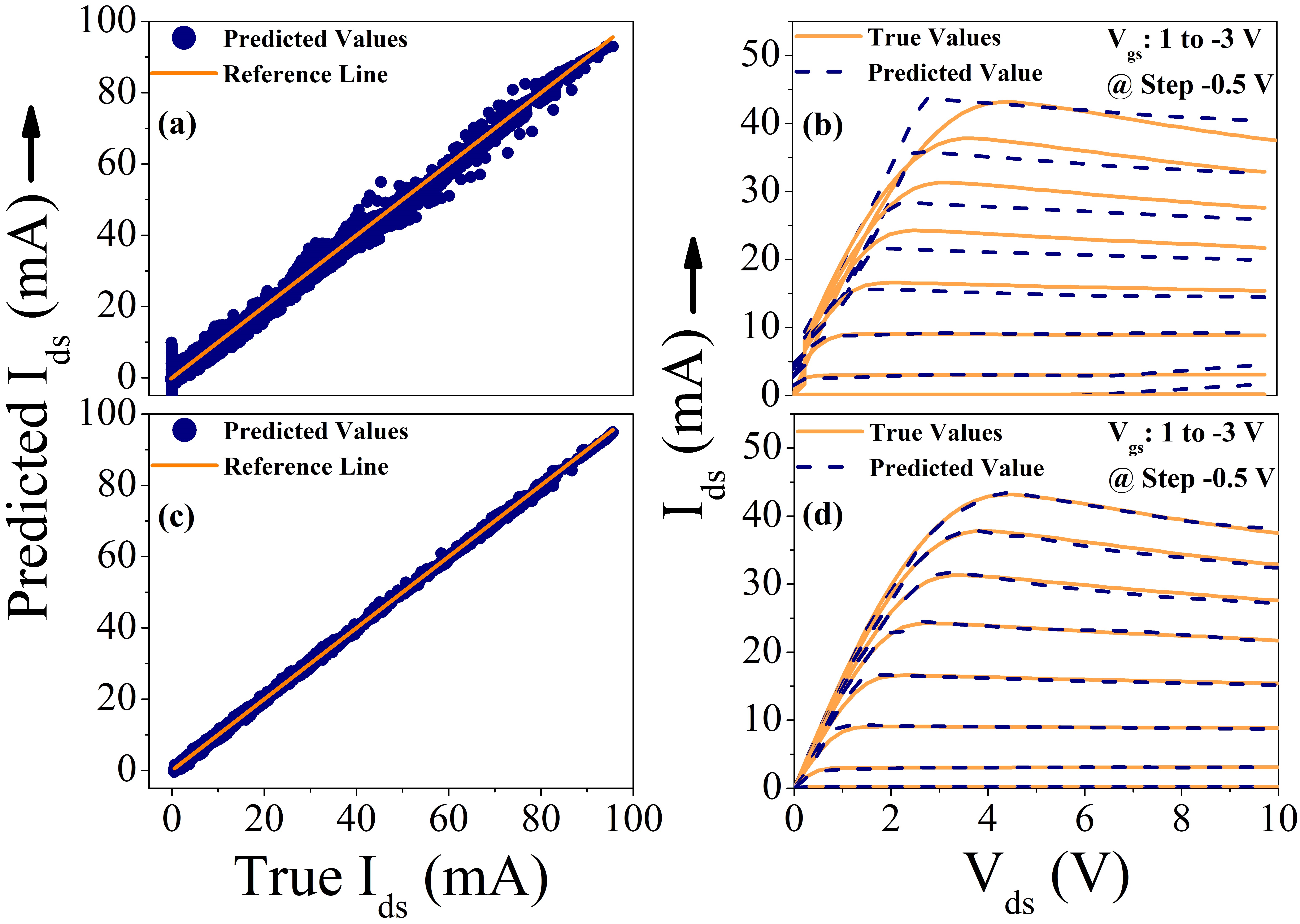}
\caption{True vs predicted values of the drain current (a, c) and I-V characteristics (b, d) for true and predicted values predicted by the (a, b) TNN and (c, d) SPGNN.}
\label{fig:PGNNHLVsnoPGNN}
\end{figure}

We have evaluated the efficiency of our approach by testing the performance in the second stage of training, where the DL model is fine-tuned on the experimental data of various sizes i.e., 100, 500, 1000, 1500 and 2000 datapoints. This analysis helps to determine the minimum amount of training data required to achieve the optimal accuracy. This fine-tuned model is termed as “semi-supervised physics guided neural network (SPGNN)”. The loss function that has been used to train the model in the second phase is standard mean squared error (MSE) function. In order to understand the contribution of the proposed framework, another DL model of the same architecture has been also trained on similar setup through traditional approach that does not have physics incorporated in it, referred as “traditional neural network (TNN)”. Fig. \ref{fig:PGNNHLVsnoPGNN} shows the performance comparison of the SPGNN and TNN. The SPGNN predicts drain current way better than the TNN for an unseen test dataset. Both the models are used to predict the I-V characteristics for a GaN HEMT that has $t_{AlGaN}$, UGW (W), L$_g$, and L$_{sd}$ of 22 nm, 100, 3 and 5 $\mu$m, respectively. The I-V characteristics predicted by the SPGNN is almost equal to the experimental values while the TNN struggles significantly in the linear as well as in saturation region for larger gate voltages. 

\begin{figure}[ht]
\centering
\includegraphics[width=4.2 in]{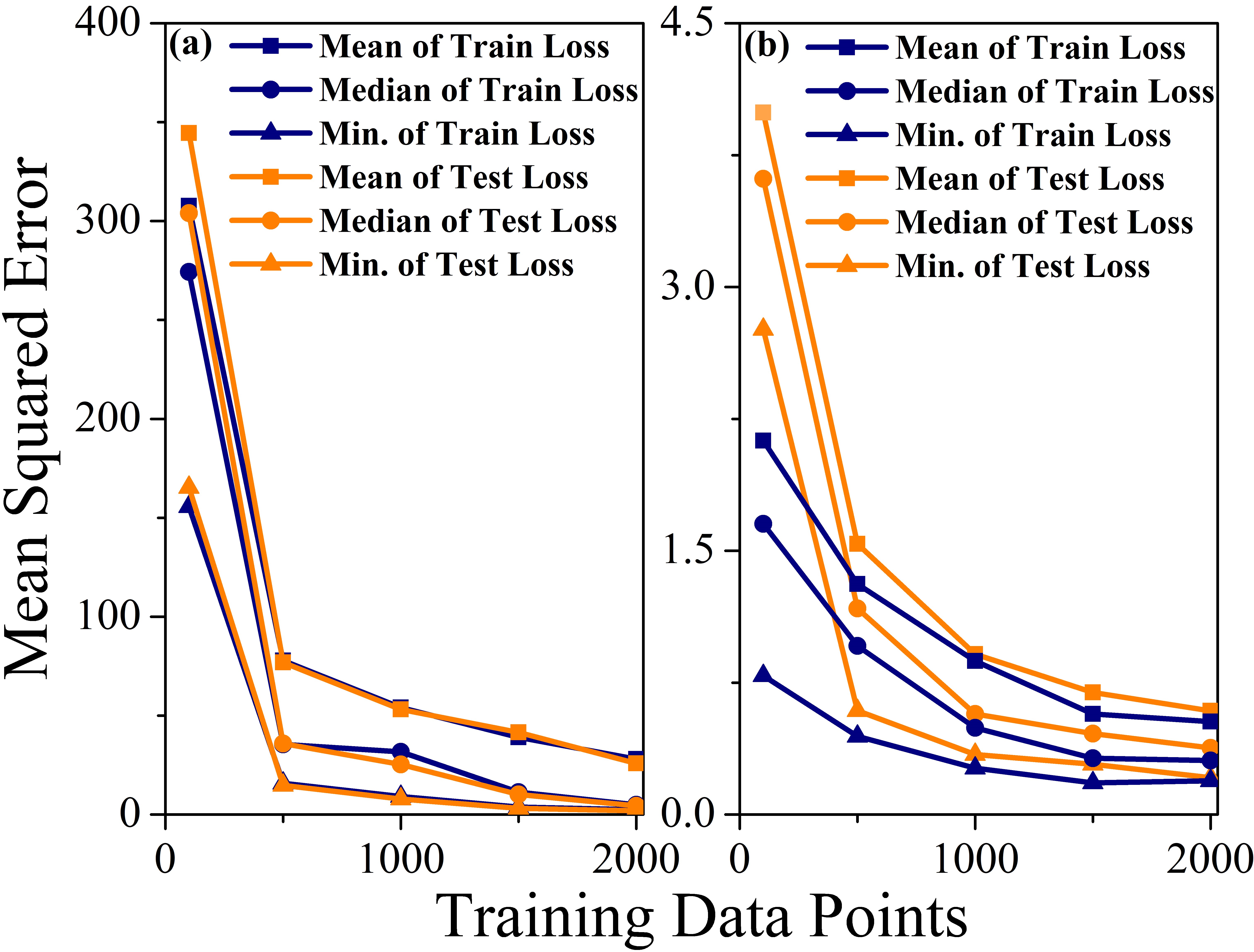}
\caption{Mean, median and minimum MSEs for training (blue color) and unseen test dataset (orange color) obtained from (a) TNN and (b) SPGNN.}
\label{fig:MSEVsDataPTs}
\end{figure}

Fig. \ref{fig:MSEVsDataPTs} shows the mean, median and minimum train and test loss of the TNN and SPGNN for various sizes of the dataset. The MSEs for the SPGNN is roughly 100 times smaller than the TNN merely by using 100 datapoints that are one sixth datapoints generated through a single GaN HEMT device (i.e., 640 datapoints). The MSEs decrease with an increase in the size of the dataset; for SPGNN it begins to saturate after 1500 datapoints. The TNN is not able to match the performance and accuracy provided by the SPGNN even when trained on 1500-2000 datapoints.The latter provides an average MSE around 60 times less than the former. The SPGNN provides better performance than the TNN despite being trained with a smaller dataset, reducing the data requirement by more than 80\%.

\begin{figure}[ht]
\centering
\includegraphics[width=4 in]{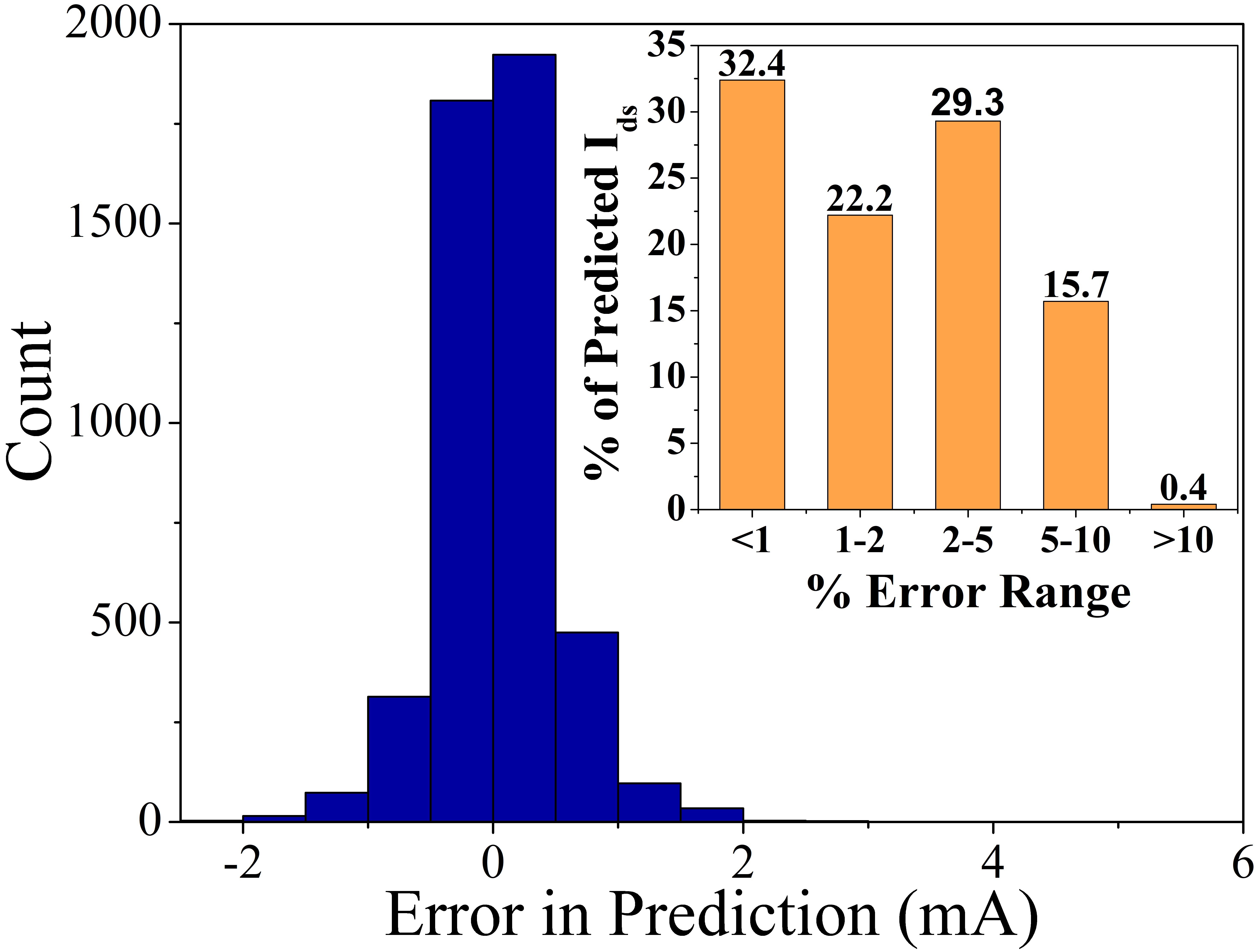}
\caption{Absolute and relative (inset) error histogram distribution for the predicted values of the drain current on the test dataset.}
\label{fig:ErrorInPrediction}
\end{figure}

Fig. \ref{fig:ErrorInPrediction} shows the error in the predicted drain currents by the SPGNN for the unseen test dataset. The model predicts almost all the values with an error less than 0.5 mA. An inset in the fig. 5 shows the percentage of the test dataset that has been predicted with a specific percentage error range. It can be seen that 32.4 \% of the data has been predicted with less than 1\% of error and only 0.4\% of the test data has been predicted with more than 10\% of error. It shows how well the SPGNN is able to predict the drain current even for the unseen data which signifies the high generalizability of the proposed DLF.

In conclusion, we have designed and developed a novel semi-supervised physics based framework for training the DL model to predict the I-V characteristics of the GAN HEMTs. The scope of the proposed framework is not only limited to the semiconductor but also to other field of research where phenomena are known. The DLF provides solutions to two major hurdles for adopting deep learning in physics domain i.e., physics consistent results and requirement of huge amount of training data. The SPGNN has been compared with the TNN to understand the importance of the proposed technique. Only 0.4\% of the predicted drain current from the SPGNN has larger error than 10\%. The developed model significantly reduces the computational resources and development time. The reported work can be very useful to the research community as it provides a better way to train the DL model even with a very small size of training dataset.\\

\section*{Acknowledgement}
The authors are thankful to the director CSIR-CEERI for his support. The authors gratefully acknowledge the support of CSIR – Nano Biosensor and Microfluidics mission mode project, Budget Head - HCP0012.

\section*{Data Availability Statement}
The implemented code will be publicly available. The data that support the findings of this study and the dataset are available from the corresponding author upon reasonable request.







\end{document}